# Fabrication of porous anodic alumina (PAA) templates with straight pores and with hierarchical structures through exponential voltage decrease technique

Leandro Sacco[1,*], Ileana Florea [1] and Costel-Sorin Cojocaru[1]

*leandro.sacco@polytechnique.edu

[1]LPICM, Ecole Polytechnique, CNRS , Université Paris Saclay, 91128 Palaiseau, France

The oxide barrier layer at the bottom of the pores has been successfully thinned by applying an exponential voltage decrease process followed by a wet chemical etching. The impact of the potential drop on the porous anodic alumina (PAA) structure has been deeply investigated, as well as the electrolyte temperature, the number of potential steps and the exponential decay rate. The results presented herein evidence that straight pores can be obtained and simultaneously remove the dielectric layer in spite of applying the exponential voltage decay during the PAA synthesis, through a smart adjustment between the anodization conditions and exponential voltage decay parameters. Additionally, the PAA structure can be tuned to fabricate hierarchically nanoporous templates with secondary pores ranging from 2 up to 10 branches. The presented simple procedure aims to become a standard step for the fabrication of the next generation PAA templates based devices.

**Introduction**

Porous anodic alumina (PAA) has attracted a considerable interest as a template for the nanostructures growth [1–7]. Such template enables fabricating self-ordered hexagonal lattice with tunable geometric features by accurately adjusting the parameters involved in the anodization process. For instance, one can synthesize PAA templates with pore diameters ranging from 10 nm up to 500 nm [8,9], high pore densities in the range $10^8$–$10^{11}$ cm$^{-2}$ and pore depths from a few hundred nanometers to the millimeter scale [10]. Regarding the synthesis of nanostructures where electrodeposition processes are involved, a critical structural parameter is the thickness of the oxide barrier layer at the pore's bottom. For example, when a thick barrier layer is perceived a high potential is required for the electrons to tunnel through such an insulating layer. Different methods have been developed in order to thin this oxide barrier layer at the bottom of the pores. For instance, aluminum can be evaporated into a conductive substrate and subsequently the pore formation is carried out until the underlying conductive support is reached [11,12]. Nonetheless, the biggest constraint of this approach, relies on the fact that an important thickness of aluminum has to be deposited since the two-step anodization process implies the elimination of a thick sacrificial layer, otherwise a low pore ordering degree is achieved. Another approach uses free-standing membranes fabricated by the complete removal of the underlying aluminum followed by an evaporation [13] or sputtered step [14] of a conductive material. However, this second approach requires two extra steps: based on a deposition of a thin metal layer followed by the transfer of the alumina membrane. However, the thickness of the PAA template is limited, since the PAA membranes have to be relatively thick in order to avoid cracks. Both mentioned processes are not desirable for the high scale devices fabrication. Dry etching can be considered as a possible alternative route for removing the oxide barrier layer [15,16]. Nevertheless, this technique involves complexes and costly technologies. The simplest effective procedure consists in decreasing the exponential voltage at the end of the second anodization process [17–20] followed by a wet chemical etching. The main disadvantage related to the exponential voltage decrease process is the formation of a branched structure at the bottom of the pores. The nanostructure's synthesis within a branched PAA structure has already been addressed and well reported in the literature [19,21–25]. Nevertheless, in all these cases, the voltage decrease is implemented to generate branched nano-scale architectures.

The oxide barrier is determined by the anodization conditions, and is generally accepted that its thickness is proportional to the applied potential. For instance, the anodization ratio (*AR)* express the dependence between the barrier layer thickness and the applied potential, defined as *AR=* $t_b/U$ where $t_b$ is the oxide barrier layer thickness and *U* is the anodization potential. There is a huge controversy concerning the role of the electrolyte temperature on the oxide barrier thickness. Some groups report that an increasing electrolyte temperature leads a lowering of the barrier layer thickness [26,27], on the other hand, there are studies which sustain that the barrier layer thickness is independent of the electrolyte temperature, and is only determined by the applied potential [28,29]. A very interesting approach is proposed by Aerts et al. [30] who compared the influence of the aluminum temperature and the electrolyte temperature on the anodization process. They conclude that the porous oxide layer displays a larger susceptibility to the electrode temperature than to the electrolyte temperature.

In this context, the main benefit of the conducted work is to highlight that the exponential voltage decrease process can be successfully implemented to thin the oxide barrier layer at the bottom of the PAA pores avoiding a branch multiplication phenomenon, or controlling the barrier thinning parameters to precisely tailor the number of branches of the hierarchically nanotemplates. The strategy adopted for reducing the branched structure relies on a conbination of an electrolyte temperature increment with a fast voltage drop. At every potential fall, the system enters into a new equilibrium state with a correlated pore cell structure which depends on the anodization parameters. A drastic change of the applied potential can stop the anodization because the system cannot be self-adjusted into the newly imposed anodization conditions. Therefore the higher electrolyte temperature implies higher density currents, consequently, for every potential drop, the system has more margins to self-adjust.

In order to estimate the number of branched, we utilize the method previously introduced [Electrical and Morphological Behavior of Carbon NanoTubes synthetized within Porous Anodic Alumina templates], where we used electrodeposited NP that reflects the fingerprint of the previous shape/state of the bottom of the pores in order to extrapolate the number of created branched by primary pore (NBPP). To accomplish the desirable PAA structure, the impact of different synthesis parameters such as the electrolyte temperature, the voltage decay rate and the number of potential drops onto the PAA structure and the NP electrodeposited was taken into account. Focus is set on the use of the focused ion beam (FIB) technique [31,32] combined with advanced 2D TEM based techniques implemented in both the TEM and STEM imaging modes of an transmission electron microscope [33,34] to get more insight about the PAA porous structure characteristics in a nanometer range, but also to determine the precise thickness of the barrier layer at the bottom pores. Such studies set the basis to build a direct relationship between the PAA architecture in terms of pores morphology, depending on the anodization conditions, and simultaneously provide valuable information for the template fabrication.

**Experimental**

**Synthesis of the PAA template**

Commercial high purity aluminum foil (Alfa Aesar 99.99%) have been cut into 1cm x 1cm small pieces and cleaned in acetone and isopropanol both in an ultra-sonic bath for 15 minutes and subsequently rinsed with deionized water. Then, the aluminum substrates have been electropolished at 5°C and 20 V during 6 minutes in a 1:4 $HClO_4:C_2H_5OH$ mixture solution. PAA templates have been fabricated by a two-step anodization process in a two-electrode electrochemical cell system. Graphite electrode has been used as a cathode and the polished aluminum foils as an anode. The first anodization has been performed during 2 hours using as electrolyte a 0.3M oxalic acid solution at various temperatures. The resulting PAA layer has been removed in a mixture solution 0.2 M $CrO_3$ and 0.6 M $H_3PO_4$ at 60°C during 2 hours in order to create nano-imprints with close-packed hexagonal arrays on the aluminum surface. The second anodization process has been executed applying same conditions than the first step, with anodization duration of 8 minutes. Immediately an exponential voltage decay process has been applied in order to thin the oxide barrier layer. The potential as function of time $U(t)$ is expressed in the following equation:

$$U(t) = V_0 exp^{(-\eta*t)} + V' \quad (1).$$

In the equation, the time constant $\eta$ gives the rate of the voltage decrease. Different $\eta$ associated at electrolyte temperatures have been used in order to study the branched multiplication phenomenon. Such parameters are summarized in table T1 in supplementary information (SI). The $V_0$ and $V'$ are free parameters whose ideal values depend on the selected constant voltage of the PAA fabrication procedure. Herein, as previous works [17], for anodization step the $V_0$ and $V'$ have been set at 50 V and -10 V respectively. The final anodization applied potential has been set at 4.97 V. Three different numbers of total voltage drops have been applied: 394, 197 and 98. Keithley 4200 power supply has been used for the anodization and the exponential voltage decrease processes. In order to fully remove the oxide barrier layer at the bottom of the pores, the resulting PAA templates have been immersed in a wet chemical etching solution 0.3M $H_3PO_4$ at 30°C during 25 minutes.

**Catalyst electrodeposition**

Nickel (Ni) nanoparticles have been deposited within the PAA template using a three-electrode configuration with an Ag/AgCl electrode used as a reference electrode (RE), the PAA template used as a working electrode (WE) and a graphite electrode which acts as a counter electrode (CE). The setup was supplied through a Bio-Logic potentiostat and in order to set the electrodeposition parameters the EC-lab software has been employed. This software allows designing the shape of the pulsed applied on the working electrode. The electrodeposition process has been performed in a Watts bath solution consisting of a mixture of 330g/L $NiO_4 \bullet 6H_2O$, 45g/L $NiCl_2 \bullet 6H_2O$, and 45g/L $H_3BO_3$. For our deposition experiment, we chose to use the pulsed mode during which the deposition step is alternated by applying a -5.5 V voltage for 5 ms between the WE and the RE, followed by 90 ms resting time in the open voltage circuit mode. For the fabrication of the entire sets of samples, the number of total sweeps has been set at 50, for all the fabricated samples, the Watts bath temperature has been kept at 20°C for all the deposition processes.

**Characterization techniques**

The morphological characterizations have been performed using a Field-Emission Scanning Electronic Microscopy (FE-SEM, HITACHI S4800). The main pore characteristics of the PAA templates, such as the pore area, the pore diameter and their boundary have been measured directly from the SEM micrographs using image treatment through the open-source "ImageJ" software [35] as in our previous work[36]. For a more detailed description of the PAA features see the SI. For the TEM analyses of all the synthesized samples, cross-section lamella preparation was necessary. Such preparation has been performed using the Focus Ion Beam technique available within a DualBeam microscope. Electron microscopy observations were carried out using two FEI-Titan electron microscopes operating at 200 kV and 300 kV. The electron microscope operating at 200kV equipped with a Cs probe corrector and a Chemistem Super-X detector allowed us achieving chemical analyses using Energy Dispersive X-ray spectroscopy (EDX).

**Results and discussion**

**PAA dependence with the anodization temperature**

In order to study the pore's bottom zone, PAA templates have been fabricated by considering different electrolyte temperatures (without applying a thinning barrier step process). The TEM micrographs obtained for each corresponding PAA templates given in figure 1(a) - (g), evidence the effect of the electrolyte temperature on the PAA template characteristics in terms of pores length and structure. In the figure is well evidenced that as the current density is affected by the electrolyte temperature (see figure S1 for more details), the PAA layer grows as the temperature increase. It is well known that the electrolyte solution and the solid electrode are characterized by different types of conductivity: electron current passes through aluminum, and ionic current through the electrolyte solution. For the current to pass through the interface between them, the following electrochemical reaction must occur: Al + 2$H_2$O --> Al(O)OH + 3H+ + 3$e^-$. In this reaction, protons are generated along with the electrons for the external circuit, which decreases the pH value at pore bottoms and promotes the in-situ dissolution of alumina leading to a rise of the anodization current [37], the ion current is driving force for oxide growth, while electronic current results in oxygen evolution under the contaminated anion layer [38]. Thus, the pore length is changing considerably from 900 nm for a low temperature (10°C) to 5 µm when a high electrolyte temperature (40°C) is used. This set of experiments also shed light on the dependence of the oxide barrier thickness as a function of the anodization temperature. Independently of the electrolyte temperature, the dielectric layer thickness remains 50 nm. Figure 1(h) shows a plot of the PAA length and the oxide barrier layer thickness as a function of the electrolyte temperature obtained by analyzing different areas of the synthesized samples. In the figure is well resumed that as the current density is affected by the electrolyte temperature, the PAA layer grows as the temperature increase. On the other hand, the barrier layer thickness not varies in the whole temperature range. Particularly, herein *AR= (1.20 ± 0.05)* nm/V which is in perfect agreement with previous works [4]. The results obtained allow us to sustain that the electrolyte temperature doesn't entail any structural advantage to thin the dielectric barrier thickness.

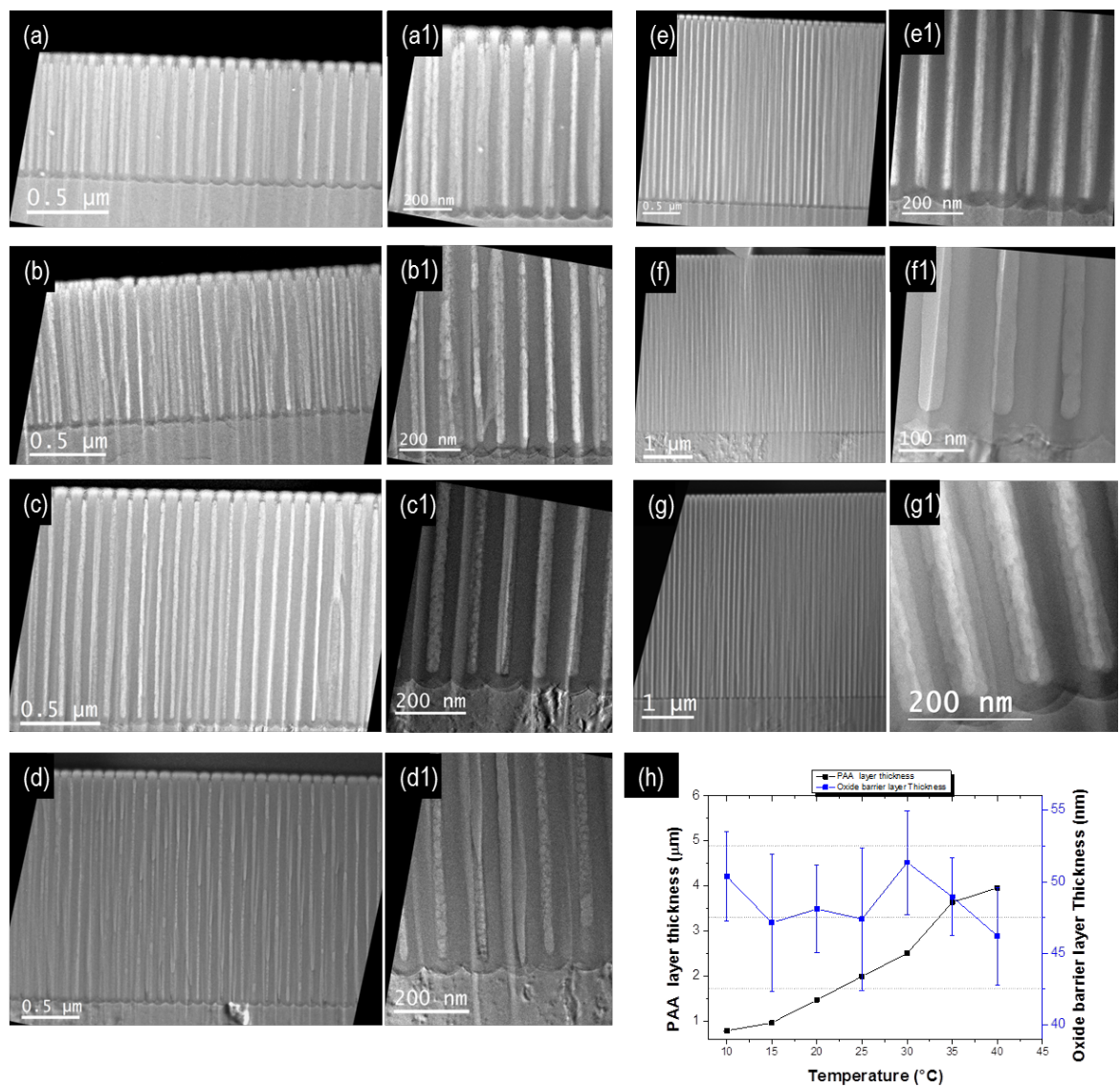

Figure1: Representative TEM micrographs of the PAA template synthesized for different electrolyte temperatures: 10°C (a), 15°C(b), 20°C (c), 25°C(d), 30°C(e), 35°C(f), 40°C (g) evidencing the PAA pores length evolution as a function of electrolyte temperature. Figure (a1), (b1), (c1), (d1), (e1), (f1), (g1) show a zoom on the bottom part of the PAA pore evidencing the presence of the 50 nm thick barrier oxide layer; (h) Graph illustrating the evolution of the PAA and oxide barrier layer thicknesses as a function of the anodization temperature.

**Exponential voltage decrease**

The branch multiplication is a phenomenon consequence of the exponential voltage decrease. In the present work, various voltage decays rate have been used with an associated electrolyte temperature. However, it's very complicated and time-consuming to establish by direct measurements, i.e. FIB manipulation and TEM observations [39], if the exponential voltage decays rate η imposed at the end of the second anodization is effective to thin the oxide barrier layer. The situation is aggravated when a branched structure is present at the bottom of the pores. Therefore, to determine an effective η value, we have taken into account two criteria's. The first one refers to the fact that the anodization voltage is not interrupted during the whole exponential voltage decrease process. This implies an increase of the current density followed the voltage drop as is pointed out in figure 2.a. Such rise of the current density guaranteed the ongoing anodization process, since the system is self-adjusted to new a cell structure associated with a steady density current, nevertheless, herein the potential drops before the system reaches a steady state. The second criterion, consider that a homogenous electrodeposition process occurs subsequently to the thinning process. In order to verify this assumption nickel nanoparticles have been deposited within the PAA template at various $\eta$. Figure 2.(b) summarizes the cumulated charged during the electrodeposition process for four different electrolyte temperatures (15°C, 20°C, 25°C and 30°C). As it can be depicted three zones in the plot are delimited as a function of the homogeneity of the electrodeposition process, figure S4 in SI summarize typical NP distribution for each zone. Taking into account that for Ni nanoparticles deposition a redox reaction takes place, successful electrodeposition process occurs when the total charge is negative. Cumulated charge (*C*) values close to zero indicates that the thickness oxide barrier layer imposed a barrier potential that hinders the electrodeposition process, consequently NP are not deposited (Zone I). When the thinning process is effective only in certain spots, electrodeposition process takes place preferentially in such zones, leading to a non-homogeneous NP distribution, as is pointed out in figure 2.b, under the presented fabrication condition such kind of NP distributions are obtained for *C* values ranging from -5 mA.s to -1 mA.s (Zone II). In case of a uniform thinning oxide barrier thickness, the electrodeposition process takes places in each pore leading to uniform distribution of individual electrodeposited NP, the associated *C* values which correspond with these kind of NP distributions ranged from -22 mA.s up to -10 mA.s (Zone III). Therefore, figure 2.b provides valuable information of which exponential decay rate ($\eta$) associated with which anodization temperature should be selected to effectively reduce the oxide barrier layer at the bottom of the pores, enabling the study of the branch multiplication phenomenon generated by the exponential voltage decrease process.

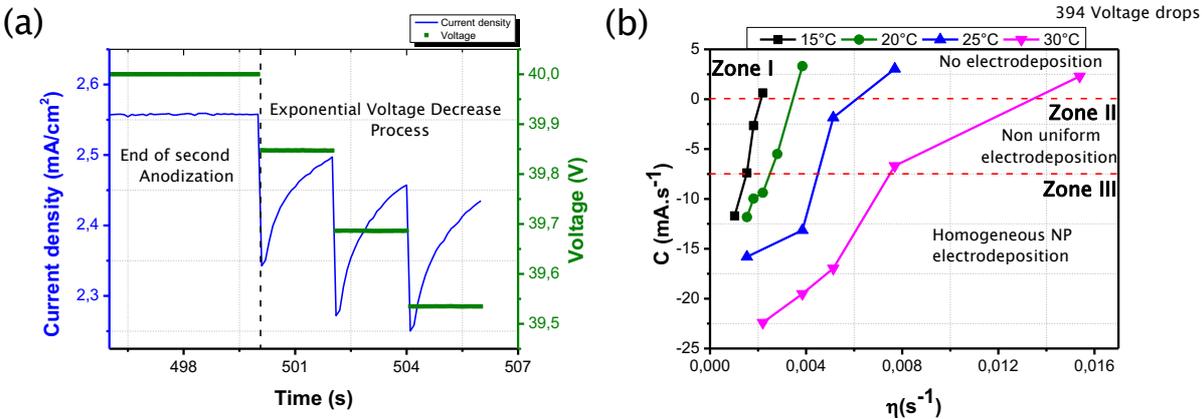

Figure 2. The two criteria considered for determining the corresponding values of the exponential voltage decays rate ($\eta$) which lead to an effective barrier thinning process. (a) Applied voltage (represented in green) and anodization current (represented in blue) curves as a function of time at the end of the second anodization process and at the beginning of the exponential voltage decrease. After every voltage drop the anodization current (blue) increase until the next voltage fall. (b) The relation between the global charges accumulated during the electrodeposition process into PAA templates fabricated at different temperatures as a function of $\eta$. Only the zone where a uniform NP distribution is obtained provides effectives $\eta$ values for the thinning process.

.

As is assess before, for many nanotechnological applications, is of paramount importance to control the number of branches or to execute the thinning barrier process avoiding a branch generation. In order to address this problem, we calculated the Number of Branches generated by Primary Pore (NBPP) as a function of the exponential voltage, applying three different total number of voltage drops. Figure 3 (a) summarizes the estimated NBPP as a function of the exponential voltage decay rate for PAA templates fabricated at different anodization temperatures by considering three different number of voltage steps (98 steps – filled triangle, 197 steps – filled circle and 398 steps-filled square). PAA templates with straight pores are pointed out in the plot by green triangles corresponding to a temperature of 32.5°C, 98 steps and the exponential voltage decay was 2.28 x$10^{-3}$s$^{-1}$. Figure 3.(b) contains a SEM micrograph illustrating the Ni NP distribution after the removal of the PAA template by a chemical etching described elsewhere [36]. The yellow circles in the micrograph point out two honeycomb patterns representative of PAA templates. In inset, the SEM image of the PAA template before the PAA removal evidencing the presence of its honeycomb pattern

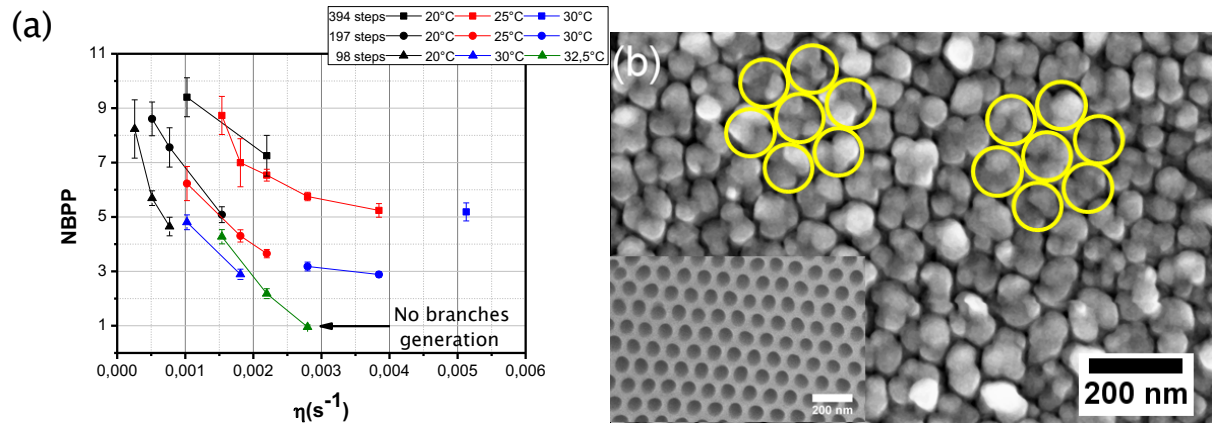

Figure 3. (a)Number of Branches generated by Primary Pore (NBPP) as a function of the exponential voltage decay rate at different anodization temperatures and various number of voltage steps. (b) SEM micrograph evidencing the Ni NP deposition after the PAA removal via a chemical etching. The PAA template has been prepared using the condition for which no branch formation is expected. The anodization temperature was set at 32.5°C, the total number of steps was 98 and the exponential voltage decay was 2.28 x$10^{-3}$s$^{-1}$. In yellow is pointed out the former pores positions.

In order to verify simultaneously the consistency of hypothesis advanced above, regarding our method based on the comparison between the number of NP and the pore densities, with the fact that straight pores can be fabricated although an exponential voltage decrease process is considered, STEM-EDX analyses have been performed. This type of analyses enables accessing direct information regarding, both the PAA pore length and the Ni NP characteristics, such as: their size and exact localization. In the STEM-HAADF micrograph represented in figure 4.(a) are shown the straight pores along the PAA layer with the Ni NP deposited at the bottom of the pores. The corresponding STEM-HAADF-EDX relative map is given in figure 4(b), obtained by considering as an elements of interests: aluminum (K$\alpha$=1.46eV), oxygen (K$\alpha$=0.523eV), and nickel (K$\alpha$=7.477eV). The relative map reveals us the presence of the Ni NP within the PAA porous structure. A closer analysis on the Ni NP present inside the pores, see the inset in figure 5. (b), evidences the presence of a 5 nm oxygen layer at the bottom part of the Ni nanoparticle, characteristic of the native oxide barrier layer of the aluminum. These findings highlight that the applied procedure successfully reduces the thickness of the oxide dielectric barrier of the PAA templates. For the very first time are shown straight pores after the application of an exponential voltage decrease process.

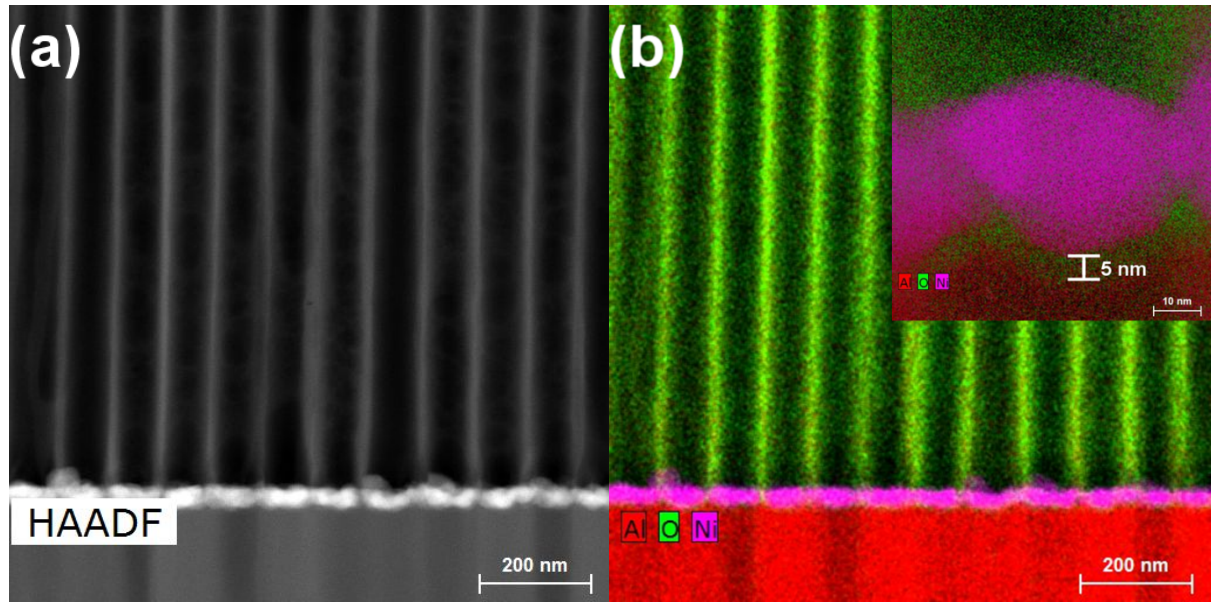

Figure 4. (a) STEM HAADF micrograph of a FIB cross-section lamella of Ni NP within the PAA template fabricated using the condition for which no branch formation is expected.(b) the corresponding STEM-EDX relative map with the aluminum represented in red, oxygen in green and nickel in pink; in inset, high-resolution STEM-EDX relative map on the Ni NP at the bottom part of the PAA pores evidencing the presence of 5nm oxide barrier layer left after the thinning procedure.

Figure 3.a provides valuable information for the fabrication of nanoporous templates with straight pores and hierarchical structures with up to 10 branches created by the primary pore. The number of total voltage steps has a great impact on the pore bottom structure. As described above, in a steady anodization the thickness of the dielectric barrier $t_B$ is proportional to the applied potential. However in the case of the exponential voltage decrease where a non-steady process takes place the barrier thickness depend on the anodized potential given by the equation (1), where is parameterized by $\eta$. Since the effectiveness of the exponential voltage decrease process is affected by the electrolyte temperature $T_e$, the oxide barrier layer also strongly depend of the electrolyte temperature. Therefore for the non-steady anodization processes, we can establish that $t_B=t_B(U,\eta,T_e)$. On the other hand, we infer that the number of voltage steps determines the NBPP. For instance, higher branches densities are obtained for PAA templates fabricate with the same $\eta$ value but with a higher number of voltage drops. This is consistent with the fact that more potentials drops entails longer voltage decrease process, consequently, the system has more time to create branches. Figure 5 summarizes the effects of the exponential voltage duration. Figure 5.(a) plots the NBPP as a function of the exponential voltage decrease duration. Such graph reflects the fact that the number of branches is approximately the same for PAA fabricated with different $\eta$ and number of total steps, but with the same anodization time. In addition, such plot evidenced that higher exponential voltage decrease durations lead to an increment of the branch multiplication phenomenon which implies a higher pore density. Then the sizes of each secondary created branch have a smaller size for longer anodization. Figure 5(b) shows the NP size as a function of the NBBP, as expected when no branches are created the pore size and the NP size matches. As more branches are created the pores reduced their size.

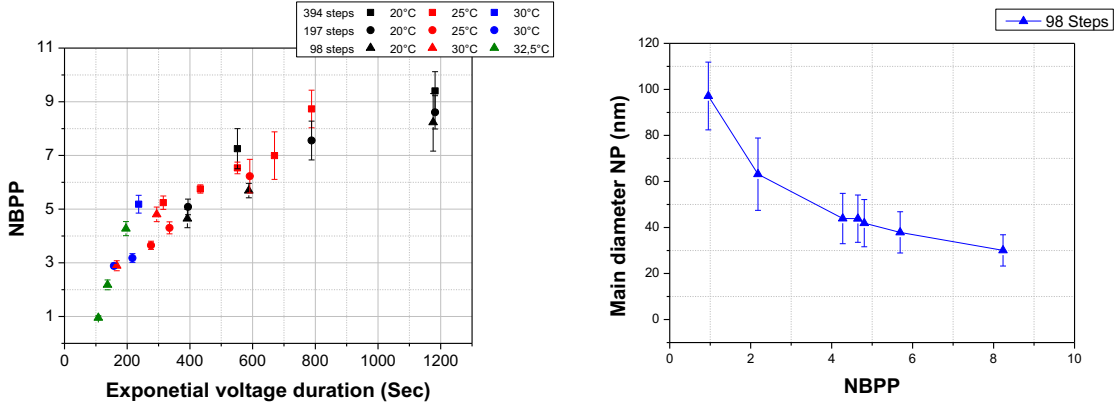

Figure 5. Influence of the exponential voltage duration. (a) NBPP as a function of the exponential voltage duration at different anodization temperatures and various number of voltage steps. (b) Nanoparticles diameter as a function of NBPP.

**Straight pore formation**

The previous results reveal that the anodization temperature and the exponential voltage decay rate mainly determine the pore structure close to oxide/aluminum interphase. In order to get a deeper understanding of the branched multiplication is interesting to roughly estimate the density of pores during the exponential decay process, and the influence of the anodization temperature on the PAA geometrical features.

The pore density ($n$) and porosity (α) at fixed potential for a hexagonal close packed array is given respectively by equations (2) and (3) [40].

$$n = \frac{2}{D_P} \sqrt{\frac{\alpha}{\pi}} \times 10^7 \quad (2)$$

$$\alpha = \frac{\pi}{2\sqrt{3}} \left(\frac{D_I}{D_P}\right)^2 \quad (3)$$

Where $D_I$ and $D_P$ are the interpore distance and the pore diameter, is generally accepted that $D_P$ and $D_I$ can be expressed as a function of the anodization potentials as in equations (5) and (6), where λ_p and λ_I are proportionality constants.

$$D_P = \lambda_P * U_{Anod} \quad (4)$$

$$D_I = \lambda_I * U_{Anod} \quad (5)$$

Manipulating these set of equations with (1), is possible to infer a temporal evolution of the amount of branches, comparing the pore density at time $n(t)$ with the initial pore density, $n(t=0)$, as express equation (6),

$$\frac{n(t)}{n(t=0)} = \frac{V_0 + V'}{V_0 exp(-\eta * t) + V'} \quad (6)$$

It is important to highlight that such relation loses their validity as the anodized PAA layers start to differ from hexagonal patterns, therefore we cannot exactly calculate the NBBP for the PAA fabricated in our work, since the final applied potential is 4.97V, which highly differs from ideally optimal anodization potential in oxalic acidic environment which corresponds to 40V[41]. However, this calculation evidenced the fact that fasters exponential decrease processes reduce the number of branches. In order to explain the formation of straights pores, we propose a pore formation mechanism which considers that the branched pores are still formed after each potential drops, but for fast voltages decay rates which effectively reduced the oxide layer, the small pore branches are dissolved due to a pore merging triggered by the high electrolyte temperature that etch the pores walls. Such idea is in harmony with the fact that the pore diameter considerably increase as the electrolyte temperature rise while the interpore distance does not significantly varies as function of the electrolyte temperature (see figure S3). Nevertheless, to achieve this PAA formation condition, a fine tune of $\eta$ is required, as evidence figure 3.a, an increment of the decay rate lead to a branch multiplication phenomenon.

## Conclusions

In summary, herein we developed a powerful procedure that enables to homogeneously remove the oxide barrier layer at the bottom of the pores which limits and complicates the synthesis of nanostructures for the development of further applications. For the first time, we directly showed the possibility to implement the exponential voltage decrease process without the creation of secondary pores. The results obtained on the described methodology clearly show that by accurately adjusting the parameters involved in the anodization process, it is possible to fabricate PAA templates nanotemplates with straight pores either hierarchically structures with up to 10 branches created per primary pore. We corroborate that in the steady anodization process the oxide barrier layer is only determined by the applied potential. However, the electrolyte temperature in the exponential voltage decrease process plays a key role since enables to increase the anodization current. Thus, faster voltage drops can be applied without interrupting the pore formation and consequently enlarging the range of $\eta$ effectives to thin the dielectric barrier layer. The number of branches created by primary pore (NBPP) is mainly determinate by such effective's $\eta$ values, and the exponential voltage duration. Shorter thinning processes duration lead to a reduction of branch multiplication phenomenon, straight pores can be fabricated when interpore walls are dissolved and the small branched pores are merged due to high electrolyte temperature that wider the pore diameter. The methodology introduced can be rapidly converted as a building block for the next generation of devices based on PAA templates.

**Acknowledgments**

L. S. gratefully acknowledges financial support from the Chaire de Recherche PSA AC3M sponsored by Citroën at the Ecole Polytechnique and Chaire EXXI sponsored by EDF at Ecole Polytechnique. The authors acknowledge financial support from the French state managed by the National Research Agency under the Investments for the Future program under the reference ANR-10-EQPX-50, pole Tempos-NanoTEM and pole TEMPOS NanoMax.

Graphical abstract

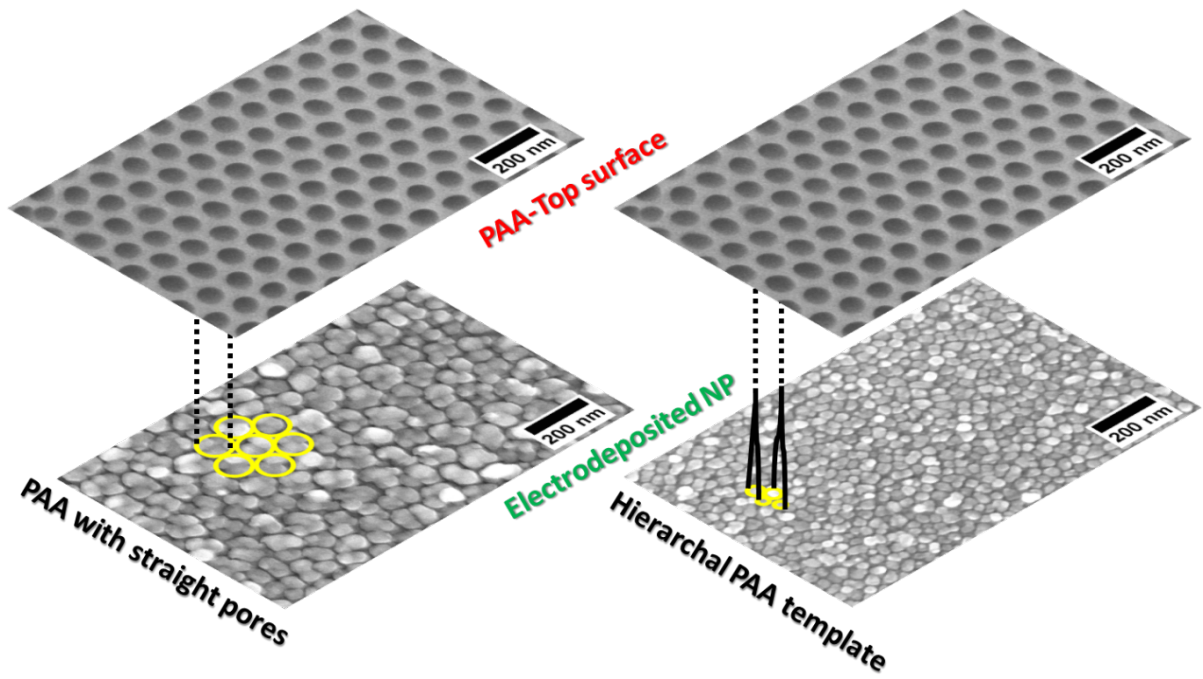

Fabrication of PAA templates with straights pores or with hierarchal structures, tailoring the number of branches at the bottom of the pores, by adjusting the parameters involved on the exponential voltage decrease process, originally applied to thin the oxide barrier layer characteristically of the PAA.